%% file: aanda.tex
\documentclass{aa}  

\usepackage{graphicx}
\usepackage{txfonts}

\usepackage{natbib}
\usepackage{journal_names}

\usepackage{graphicx}	
\usepackage{amsmath}	
\usepackage{amssymb}	
\usepackage{booktabs}

\usepackage{color}
\usepackage{xcolor}
\usepackage{longtable}

\newcommand{\Potter}{\textsc{Paper I}}
\newcommand{\Potters}{\textsc{Paper I }}
\usepackage{dirtytalk}
\usepackage[nolist,nohyperlinks]{acronym}
\usepackage[normalem]{ulem}

\usepackage{soul}

\begin{document}

   \title{High-speed photometry of the eclipsing polar UZ Fornacis}
   
   \titlerunning{High-speed photometry of the eclipsing polar UZ Fornacis} 

   \author{Z. N. Khangale
          \inst{1,2}, S. B. Potter \inst{1}, E. J. Kotze \inst{1}, P. A. Woudt \inst{2}
          \and
          H. Breytenbach \inst{1,2} 
          }

   \institute{South African Astronomical Observatory (SAAO), PO Box 9, Observatory 7935, Cape Town, South Africa\\
              \email{khangalezn@saao.ac.za}
         \and
             Department of Astronomy, University of Cape Town, Private Bag X3, Rondebosch 7701, South Africa\\
             }
     \authorrunning{Z. N. Khangale et al.}

   \date{Received 09 August 2018; accepted 21 November 2018}

\abstract{We present 33 new mid-eclipse times spanning approximately eight years of the eclipsing polar UZ Fornacis. We have used our new observations to test the two-planet model previously proposed to explain the variations in its eclipse times measured over the past $\sim$35 years. We find that the proposed model does indeed follow the general trend of the new eclipse times, however, there are significant departures. In order to accommodate the new eclipse times, the two-planet model requires that one or both of the planets require highly eccentric orbits, that is, $e \geq$ 0.4. Such multiple planet orbits are considered to be unstable. Whilst our new observations are consistent with two cyclic variations as previously predicted, significant residuals remain. We conclude that either additional cyclic terms, possibly associated with more planets, or other mechanisms, such as the Applegate mechanism are contributing to the eclipse time variations. Further long-term monitoring is required.}


\maketitle

\section{Introduction}

UZ Fornacis (hereafter UZ For) is an AM Herculis-type eclipsing magnetic cataclysmic variable (CV) star discovered with the EXOSAT as an X-ray source, EXO 033319-2554.2 \citep{1987IAUC.4486....1G}.  
It has an orbital period of $\sim$126.5 min and spectral type of M4.5 \citep{1991MNRAS.253...27B}. 
UZ For has been studied extensively on a wide range of wavelengths including optical \citep[e.g.][]{1989ApJ...347..426A,1990A&A...230..120S,1991MNRAS.253...27B}, x-ray \citep[e.g.][and references therein]{2001ApJ...562L..71S}, ultra-violet \citep{1996ApJ...468..883S} and extreme-ultraviolet \citep[e.g.][]{1993AAS...182.4110W,1994BAAS...26..793S,1995ApJ...445..909W}. 
UZ For displays one or two accretion spots depending on the accretion state, with magnetic fields of $\sim$53 and $\sim$48 MG \citep{1988A&A...195L..15B,1989ApJ...337..832F,1991MNRAS.253...27B}. 

The evolution of CVs is governed by the binary's angular momentum and stellar masses, and how these parameters change over time. It is generally understood that CVs evolve from longer to shorter orbital periods. After the common-envelope phase, when mass transfer begins, angular momentum is constantly being exchanged between the white dwarf (WD) and red dwarf resulting in the shrinking of the orbital separation and thus reducing the orbital period of the binary. For systems with short orbital periods ($\leq$ 3 h), the main angular momentum loss mechanism is gravitational radiation \citep{1971ApJ...170L..99F,1981ApJ...248L..27P}. At longer periods ($\geq$3 h), the angular momentum loss is driven by a mechanism called magnetic braking \citep{1981A&A...100L...7V,1983ApJ...275..713R}. In one example, \cite{2006MNRAS.365..287B} argue that the period change in the post-common-envelope binary NN Ser can be explained by either a genuine angular momentum loss from the system or the presence of unseen companions around the binary system. Their observations showed that NN Ser was losing angular momentum at a rate predicted by \cite{1983ApJ...275..713R} but only if they assume that magnetic braking was not cut off as the secondary reaches 0.3 $M_{\odot}$. However, \cite{2010A&A...513L...7S} present the results of 670 post-common-envelope binaries and found strong evidence for disrupted magnetic braking at the fully convective boundary. 

The accretion rate in CVs is highly variable, and the majority of these systems change from a high state to a low state and back over timescales ranging from days to months and years. 
These variations range from both flickering typical of CVs on time-scales of minutes, and significant changes in the overall shape of the light curves. A number of factors can contribute to those variations which include the activity and the shape of the secondary star. For example, the UZ For binary system has been reported to switch between the faint state \citep{1991MNRAS.253...27B} and the bright state \citep{1998ApJ...501..830I} on a timescale of years. 
The one constant in the light curve of eclipsing polars is the eclipse of the WD by the secondary star: generally, the ingress and egress last for $\sim$30 s. 
For UZ For the ingress and egress are rapid at $\sim$3 s enabling accurate determinations of mid-eclipse times, based on the midpoint between the ingress and egress.    

Recently, a number of systems have been found to show variations in their times of eclipse, for example, NN Ser \citep{2009ApJ...706L..96Q,2010A&A...521L..60B}, HU Aqr \citep{2011MNRAS.414L..16Q,2015MNRAS.448.1118G} and DP Leo \citep{2010ApJ...708L..66Q,2011A&A...526A..53B}. 
Neither gravitational radiation nor magnetic braking is sufficient to explain the period changes observed. 
Several explanations for these eclipse time variations have been offered in the literature which include either solar-type magnetic cycles of the secondary star (Applegate mechanism: \citealt{1992ApJ...385..621A}) or the presence of circumbinary planets in an orbit around the binary, for example \cite{2006MNRAS.365..287B}. 

According to Applegate's mechanism, the period variations result from quasi-periodic changes in the quadruple moment of the secondary star due to magnetic activity. 
In this model, it is assumed that a strong magnetic field is produced by a dynamo cycle resulting in the redistribution of the angular momentum within the star and hence a change in its quadruple moment. 
\cite{1989SSRv...50..219H} found that for Algols there is a strong connection between the orbital period variations and the presence of magnetic activity. However, \cite{2006MNRAS.365..287B} report that the orbital variation in NN Ser can not be explained by the Applegate mechanism. 

The original Applegate model linking magnetic activity to orbital period variations has been reviewed by different authors, for example \cite{1998MNRAS.296..893L}.  
Recently, \cite{2016A&A...587A..34V}, presented an improved version of Applegate's mechanism which now includes the angular momentum exchange between a finite shell and the core of the star to derive the general conditions under which the Applegate's mechanism can operate. 
They find that, out of the 16 systems that were analysed, only four systems (e.g. QS Vir, DP Leo, V471 Tau, BX Dra) could be explained by the improved Applegate's mechanism. For the remaining systems, more than the total energy generated by the secondary star is necessary to power the binary's period variations. They note that for UZ For and three other systems, the ratio of energy required to power the improved Applegate mechanism to the total energy generated by the secondary star is almost unity. In the case of NN Ser, the ratio of energy required to power the improved Applegate mechanism to the total energy generated by the secondary star is greater than unity, implying that it can not be explained by magnetic activity. 

In the case of the period changes which are due to the presence of a companion(s) in the binary, the observed minus calculated (O - C) time of eclipses vary as the binary orbits the centre of mass of the system. These small variations appear as periodic variations in the O - C diagram due to the light traveltime effect. 
However, high-eccentric and/or multi-planet solutions are required to fully explain the O - C variations (e.g. HU Aqr: \citealt{2011MNRAS.416L..11H,2012MNRAS.419.3258W}). These can be problematic for dynamically stable orbits. In addition,
\citet{2012MNRAS.420.3609H} re-analysed the eclipse times of HU Aqr and also found that the best-fitting model requires dynamically unstable solutions with high eccentricities for two companions. Nevertheless, examples of dynamically stable solutions are possible if non-coplanar, high eccentric and even retrograde orbits are used \citep{2012MNRAS.420.3609H,2015MNRAS.448.1118G}. 

On the other hand, a two-planet solution for NN Ser has been shown to be dynamically stable and survived follow-up eclipse time measurements   \citep{2012MNRAS.425..749H,2013A&A...555A.133B,2014MNRAS.437..475M}. 
Furthermore, \cite{2016MNRAS.460.3873B} reported that the long-period quadratic term in the model of NN Ser is in the direction of lengthening period and can not be explained by natural processes that lead to angular momentum loss. 
This leaves the circumbinary planet hypothesis as an option to explain the periodicities in NN Ser. This is further supported by the existence of the circumbinary disc around NN Ser \citep{2016MNRAS.459.4518H}.

Recently, \cite{2016MNRAS.460.3873B} carried out a long-term programme of eclipse measurements on 67 WDs in close binaries to detect the period variations. Their results show that systems with baselines exceeding ten years, and with companions of spectral types M5 or earlier, appear to show greater eclipse times variations than systems with companions of spectral types later than M5. They found this to be consistent with an Applegate-type mechanism. However, they also considered it reasonable to assume that some planetary systems could exist around evolved WDs binaries, for example NN Ser \citep{2013A&A...555A.133B,2016MNRAS.459.4518H}. A recent study by \citet{2018A&A...611A..48P} agrees with the earlier conclusion by \cite{2016MNRAS.460.3873B} that higher values of O - C residuals are found with secondary companions of spectral type M5/6 or possibly earlier as a result of an Applegate mechanism. The spectral type of UZ For is dM4.5 \citep{1988A&A...195L..15B} suggesting that an Applegate-type mechanism could be significant in this system.

\citet{1988A&A...195L..15B} used a quadratic ephemeris to fit eight eclipse times of UZ For which possibly indicated a decrease in orbital period. 
Follow-up studies by \cite{1994IBVS.4075....1R} and \cite{1998ApJ...501..830I} pointed at an increasing orbital period of UZ For. \cite{2001MNRAS.324..899P} derived a linear ephemeris using their three eclipses combined with earlier six timings by \cite{1991MNRAS.253...27B}. However, they noticed that their new ephemeris leaves residuals of order $\pm$50 s when compared with historical data taken earlier.  
\citet{2010MNRAS.409.1195D} presented 44 mid-eclipses of UZ For and noticed a deviation from linear and quadratic trends in the O - C diagram of UZ For. They explained the deviations by adding a sinusoidal term to the ephemeris and attributed the cyclic variation as due to a planet with the period of $\sim$23(5) years. 
\cite{2011MNRAS.416.2202P} presented new mid-eclipse times, including those from literature and included in \cite{2010MNRAS.409.1195D}, over the 28 year baseline and noticed a deviation from linear and quadratic trends with amplitudes of 60 s. 
They interpreted this as the result of two cyclic variations due to two extrasolar planets in orbit around the binary with periods of $\sim$16(3) years and $\sim$5.25(25) years. However, they did not rule out the possible effect of a magnetic cycle mechanism. 

In this paper we present new photometric observations of the eclipsing system UZ For spanning an additional eight years with the aim of investigating the purported cyclic variations (\citealt{2011MNRAS.416.2202P}, hereafter \Potter) further. 
Sect. \ref{phot} gives an account of all our observations. 
In Sect. \ref{sec:oc}, we show the new eclipse O-C results as well as updated fitting parameters for the planetary model. 
We provide a general discussion in Sect. \ref{sec:disc} and conclusion in Sect. \ref{sect:con}.

\section{Observations} \label{phot}

\input{Phot_obsev_log}

Photometric observations were made between 2011 March and 2018 February on the 1.0-m and 1.9-m telescopes located on the Sutherland site of the South African Astronomical Observatory (SAAO), using either the HIgh-speed Photo-POlarimeter (HIPPO, \citealt{2010MNRAS.402.1161P}) or the Sutherland High-speed Optical Camera (SHOC, \citealt{2011epsc.conf.1173G,2013PASP..125..976C}). The log of observations is shown in Table \ref{Table:phot}. All the observations were made in good seeing conditions. 

The HIPPO instrument was operated in its photo-polarimetry mode (all-Stokes) and the observations were clear filtered (3500--9000 \AA{}). 
Background sky measurements were taken at frequent intervals during the observations. 
All of our observations were synchronized to GPS to better than a millisecond. 
Data reduction was carried out following the procedures described in \citet{2010MNRAS.402.1161P}. A total of 16 eclipses were obtained in photometric conditions.  

The SHOC detector was used in a frame-transfer mode with a clear filter, a binning of 8 $\times$ 8 or 16 $\times$ 16 and exposure times of one second. Differential photometry was performed on the resulting data cubes using the SHOC-pipeline described in \citet{2013PASP..125..976C}.
A total of 17 high-time resolution and high signal-to-noise ratio eclipses of the target were obtained. 

A total of 33 new eclipses of UZ For were obtained.
All the eclipse times were corrected for the light travel-time to the barycentre of the solar system, converted from Julian dates to Barycentric dynamical time (TDB) system as Barycentric Julian date (BJD, \citealt{2010PASP..122..935E}). 
This was done in order to remove any timing systematics, particularly due to the unprecedented accumulation of leap seconds with universal time central (\textsc{utc}) and the effects due to the influence of Jupiter and Saturn when heliocentric corrections only are applied. 
The times of mid-eclipse were determined as the midpoint between the steep drop to minimum (ingress) and the steep rise out of minimum (egress) of the main accretion spot of UZ For as shown in Figure \ref{figure:phot}. The ingress and egress are marked with blue dashed lines and the adopted time of eclipse is marked by the black dashed line in Figure \ref{figure:phot}. 
We estimated an error of approximately 0.00003 days for each of the new eclipses.  
The new mid-eclipse times were combined with the 42 mid-eclipse times presented in \Potters to give a new total of 75 and we present the results in Table \ref{tab:01}.

\section{Results}\label{sec:oc}

\subsection{Eclipse profiles}

\begin{figure*}
\begin{center}$
\begin{array}{cc}
\includegraphics[height = 6.00cm, width = 7.5cm ]{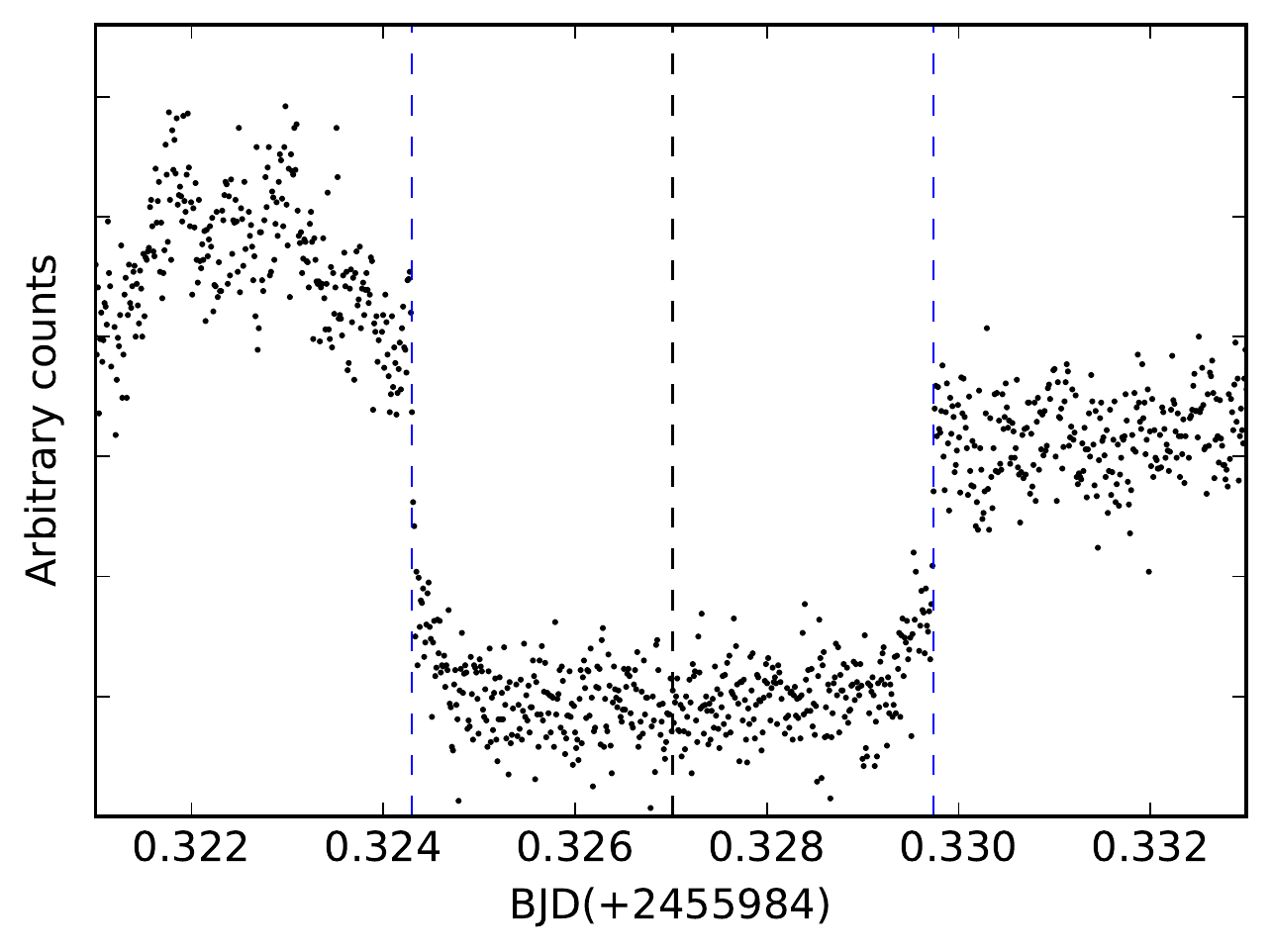}
\hspace{0.25cm}
\includegraphics[height = 6.00cm, width = 7.50cm ]{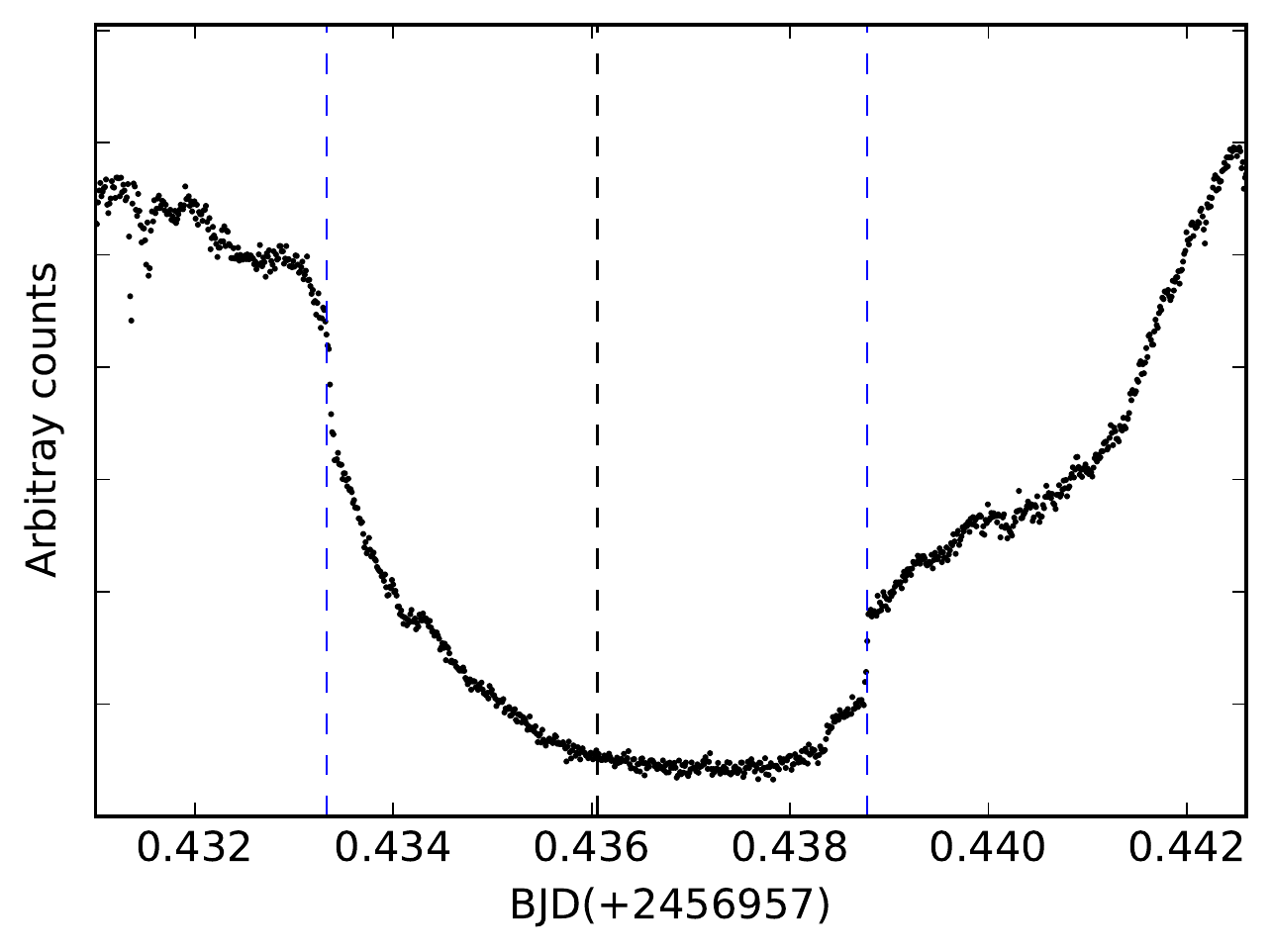}
\end{array}$
\end{center}
\caption{Two example eclipses of UZ For obtained with HIPPO (left, 26 Feb 2012) and SHOC (right, 26 October 2014) instruments. The vertical blue dashed lines 
represent the times of ingress and egress of the main accretion spot, whereas the vertical black dashed line marks the time of mid-eclipse.}
\label{figure:phot}
\end{figure*}

Figure \ref{figure:phot} shows two of the eclipse profiles of UZ For obtained during a faint and bright state (left and right panels respectively). There was no flux calibration performed, however, the relative signal-to-noise of the two datasets is indicative of the relative brightness. The eclipse profiles can be understood in the framework of the standard polar model \citep{1988A&A...195L..15B,1989ApJ...337..832F,1991MNRAS.253...27B}. UZ For shows either one- \citep{1991MNRAS.253...27B,1998ApJ...501..830I} or two-pole \citep{2001MNRAS.324..899P} accretion spots depending on the accretion states. 
Both panels show clearly defined ingresses and egresses of the main accretion spot lasting a few seconds, indicated by vertical blue dashed lines.   

The faint state eclipse profile (Fig. \ref{figure:phot}, left panel) appears flatter and noisier than the brighter state eclipse profile. 
It is apparent that there are two clear stages in both the ingress and egress. 
The ingress is defined by a fast drop ($\sim$3 s) in counts followed by a more gradual decline to a minimum ($\sim$50 s). The eclipse remains flat for ($\sim$380 s) and the egress begins with a slow initial rise ($\sim$30 s) followed by a rapid rise ($\sim$3 s). 
The shape of this eclipse profile is similar to those of the low-state of UZ For presented in \citet{1991MNRAS.253...27B}. 

The bright state eclipse profile (Fig. \ref{figure:phot}, right panel) also shows a fast drop in counts during ingress ($\sim$3 s) but now followed by a brighter and longer decline ($\sim$260 s) to a minimum, compared to the fainter eclipse profile, resulting in a shorter time for the flat part ($\sim$169 s). The egress begins with a two-step increase in counts ($\sim$34 s), possibly indicating accretion at the second pole, followed by a rapid increase ($\sim$3 s) as the main pole egresses.
 
The various stages are consistent with bright-state accretion with two accretion spots, and a brighter magnetic accretion stream contributing to the overall brightness of the system. Our bright-state eclipses are similar to those presented by \citet{1998ApJ...501..830I}. In both cases, the faint and bright state, the eclipse width remained the same at $\sim$472 s consistent with the eclipse of the main accretion spot. 

\subsection{The new eclipse times}

\begin{figure*}
\begin{center}
\includegraphics[height = 13.00cm, width = \textwidth  ]{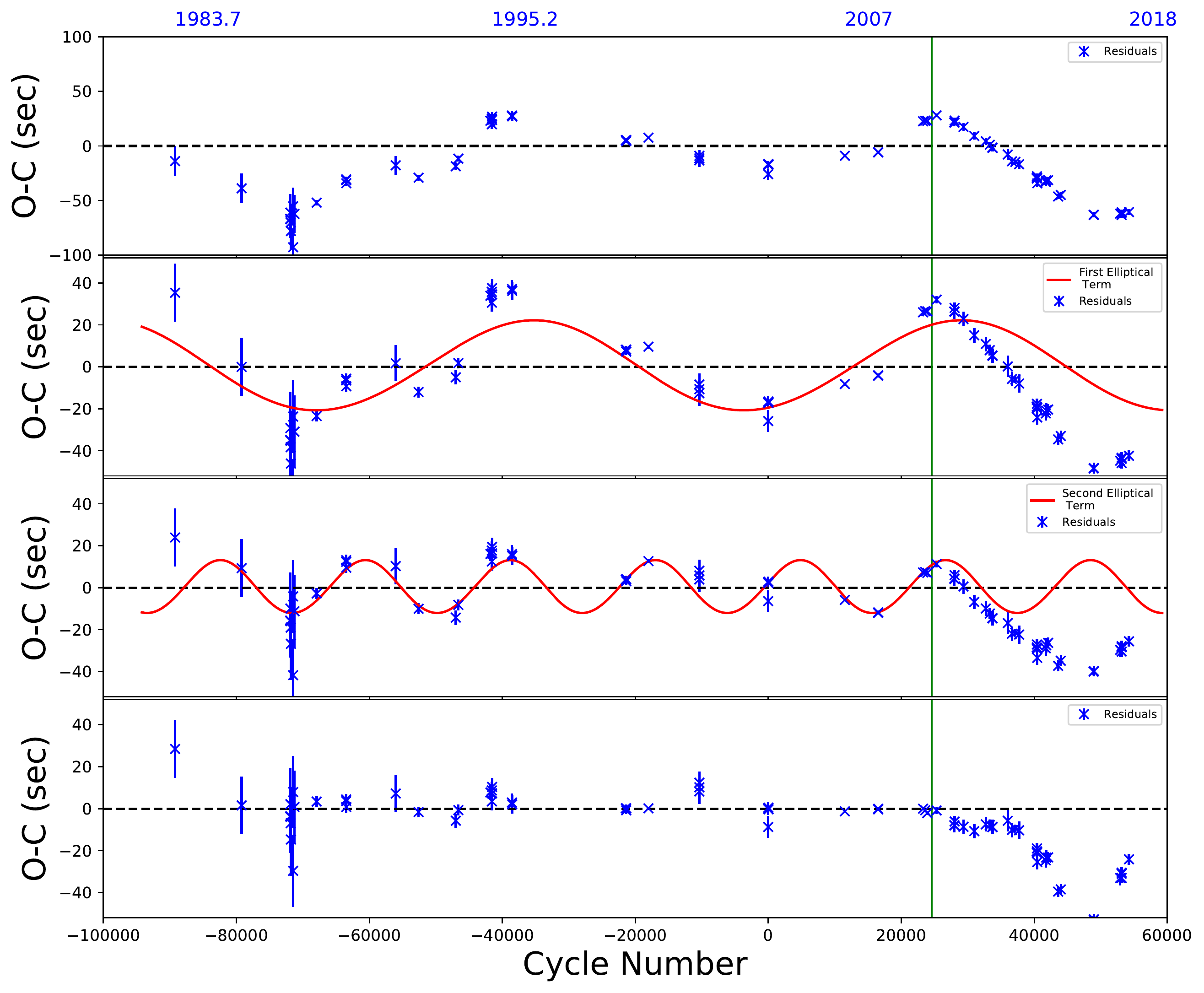}  
\end{center}
\caption{O - C diagram of UZ For from \Potters but with new eclipses added.  The vertical green lines separate the eclipse times from literature (to the left) and our new eclipse times (to the right). (See text for more details).  
}
\label{figure:oc_old}
\end{figure*}

Table \ref{tab:01} lists all of our new mid-eclipse times as well as those presented in \Potter. 
We used the epoch ($T_0$) and the orbital period from \Potters to calculate the cycle number for each of the new mid-eclipse times. 
The orbital period is accurate enough to assign cycle numbers to the entire 35 years of eclipses.  

Figure \ref{figure:oc_old} shows the O - C diagram of UZ For spanning $\sim$35 years. The crosses to the left of the vertical green line are those eclipse times presented in \Potters (up to 2010) and the crosses to the right are the new eclipse times amounting to an additional eight years. Over-plotted is the solution from \Potters (red curves). The top panel shows the O - C after subtraction of the linear ephemeris and it is well known that a linear fit does not give a valid description of the data. The residuals on the top panel appears periodic and therefore, we need more terms in addition to the quadratic term in order to describe the O - C diagram of UZ For. The second panel shows the O - C after subtraction of the quadratic term with the first elliptical term overplotted (red curve), the third panel show the O - C residuals after subtraction of the first elliptical term with the second elliptical term overplotted (red curve), and the bottom panel show the final O - C residuals after subtraction of the second elliptical term. 
The new mid-eclipse times were not included in the fit. We note that the original fit was not the formal best solution but instead included the constraint that the eccentricities of the elliptical terms were $\leq$ 0.1 in order to be consistent with a stable two-planet model. 

The mathematical function fitted to the old data (the crosses to the left of the green lines) predicted a maximum in the O - C residuals around $\sim$2010-2011 
followed by a gradual decline in the residuals (Fig. \ref{figure:oc_old}, second and third panels) followed by a minimum in the O - C residuals. 
Our new eclipse times (the crosses to the right of the green lines) agree with the predictions in the sense that a maximum in the O - C residuals occurred around $\sim$2011-2012 followed by a gradual decline in the residuals. 
However, the gradual decline appears steeper than predicted and occurs earlier and continues to diverge after $\sim$2013 until the minimum in the residuals is reached in 2016-2017. 
The minimum is followed by an upturn in the O - C residuals in agreement with predictions.

\subsection{Calculating a new O - C}

The model from \Potters and shown in  Fig. \ref{figure:oc_old} consists of a combination of a quadratic and two elliptical terms, Eq. \ref{equ.sec} below: 
\begin{equation}
\centering
\label{equ.sec}
\begin{split}
 T({\rm{BJD_{TDB}}}) = T_0 + P_{{\rm{bin}}}E + AE^2 
 \\+ K_{{\rm{bin}},(3)} {\rm{sin}}(\upsilon_3 - \varpi_3)\frac{[1 - e^2_3]}{[1 + e_3 \rm{cos}(\upsilon_3)]} 
 \\+ K_{{\rm{bin}},(4)} {\rm{sin}}(\upsilon_4 - \varpi_4)\frac{[1 - e^2_4]}{[1 + e_4 \rm{cos}(\upsilon_4)]}.
\end{split}
\end{equation}

\noindent
Substituting $\upsilon_{(3,4)}$ = $(E + T_{(3,4)})f_{(3,4)}$ in the above equation result in the following:
\begin{equation}
\centering
\label{equ.sec1}
\begin{split}
 T({\rm{BJD_{TDB}}}) = T_0 + P_{{\rm{bin}}}E + AE^2 
 \\+ K_{{\rm{bin}},(3)} {\rm{sin}}((E + T_{3})f_{3} - \varpi_3)\frac{[1 - e^2_3]}{[1 + e_3 \rm{cos}((E + T_{3})f_{3})]} 
 \\+ K_{{\rm{bin}},(4)} {\rm{sin}}((E + T_{4})f_{4} - \varpi_4)\frac{[1 - e^2_4]}{[1 + e_4 \rm{cos}((E + T_{4})f_{4})]},
\end{split}
\end{equation}

\noindent 
where $T_0$ is the time of epoch, $P_{{\rm{bin}}}$ is 
the orbital period of the binary in days, $A$ is the quadratic parameter and $E$ is the binary cycle number which comprises the quadratic term of the ephemeris. 
The remaining ten parameters were introduced in the context that the variations in the eclipse times are due to the light travel-time effect caused by the gravitational influence of a third and fourth body orbiting the central binary system. 
$K_{{\rm{bin}},(3,4)}$ are the amplitudes of the eclipse time variations as a result of the light travel-time effect of the two bodies.  
$\upsilon_{(3,4)}$ are the true anomalies of the two bodies and, where $\upsilon_{(3,4)}$ = $(E + T_{(3,4)})f_{(3,4)}$, which progresses through 2$\pi$ over the orbital periods [$P_{(3,4)}$] and are functions of $E$. 
$T_{(3,4)}$ are the times of periastron passages and $\varpi_{(3,4)}$ are the longitudes of periastron passage measured from the ascending node in the plane of the sky. Lastly,  
$e_{(3,4)}$ and $f_{(3,4)}$ are the eccentricities and orbital frequencies of the two bodies.

We next investigated using Eq. \ref{equ.sec} on all of the eclipse times. However Eq. \ref{equ.sec} has thirteen variables for minimization, the first three are associated with the quadratic term and the remaining ten are associated with the two elliptical terms. Therefore we generated a starting grid, for minimizations, of 10360 starting points for the two elliptical frequencies evenly spaced with $f_3$ (between 0.0000552 and 0.000138) and $f_4$ (between 0.0001794 and 0.0004555) cycles per binary cycle. The remaining  eight of the ten elliptical parameters were randomized to within reasonable values e.g. the eccentricities took random values between 0 and 1.  We then performed simultaneous least-square fitting for each starting set of parameters. During minimization all the parameters were allowed to vary and as expected all the minimizations did not converge to a single solution but instead gave a range of solutions with final reduced $\chi^2$  between 2.06 and 397.27. Rerunning the minimizations with a new grid of random values for the parameters give the same general results. We explore the solutions in the next two sub-sections.
  
\subsection{Distribution of the elliptical frequencies}

\begin{figure*}
\centering
\includegraphics[height = 11.00cm, width = 13.0cm  ]{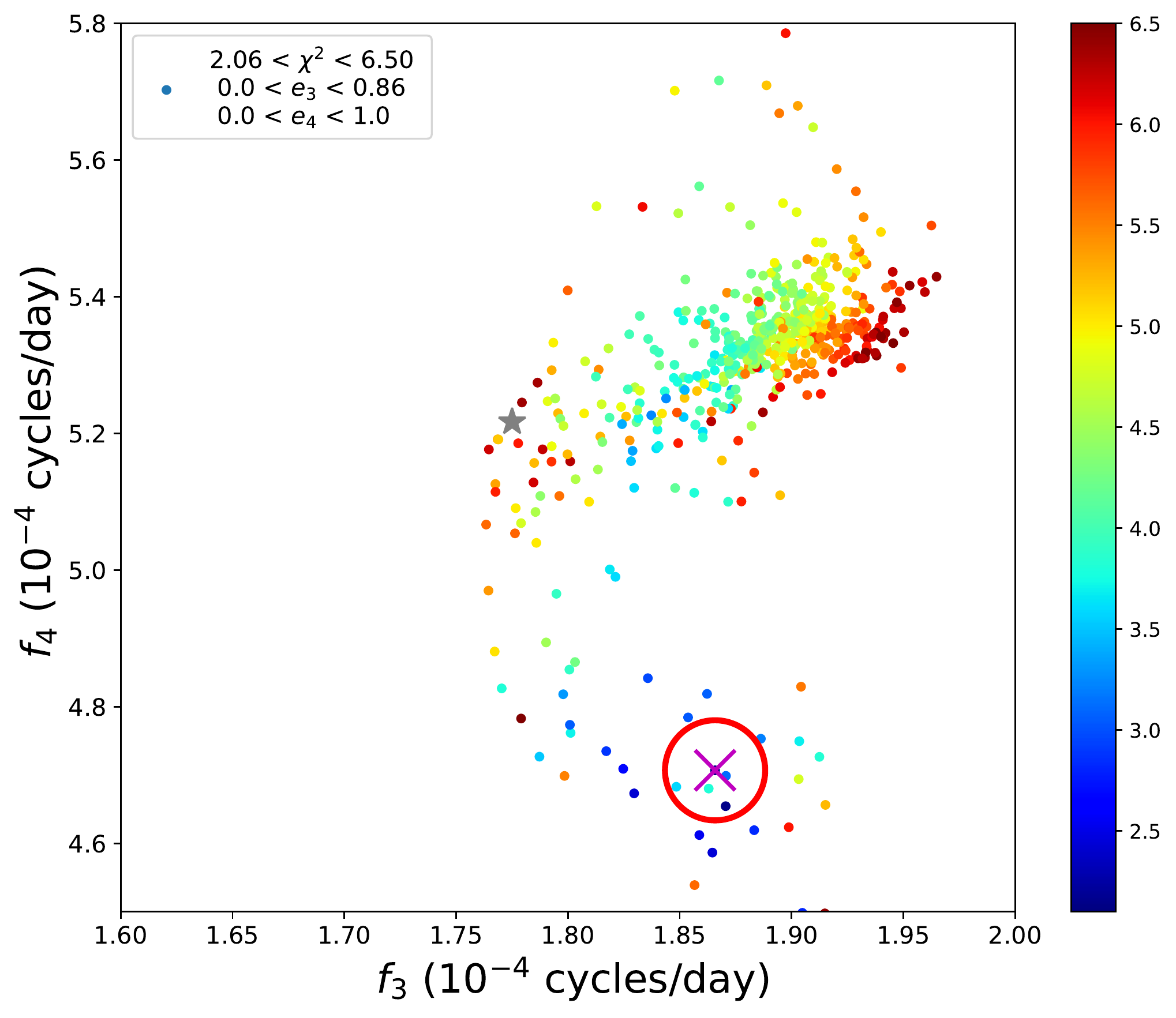}  
\caption{Reduced $\chi^2$ parameter space for the two frequencies. The points have been colour-coded with dark blue and dark red representing lower and higher reduced chi-square:  2.06 $\leq \chi^2 \leq$ 6.50. The solutions with $e \geq$ 1.0 were excluded since their solutions will give parabolic and hyperbolic orbits. The grey star marks the location of the solution adopted in \Potter, whereas the red circle and magenta cross mark the solution with the lowest reduced $\chi^2$ adopted as the best-fit for this paper. }
\label{figure:ff3}
\end{figure*}

The distribution of the reduced $\chi^2$ (not shown) forms an asymmetric Gaussian profile peaking at approximately five. 
The Gaussian profile peaks rapidly from a reduced $\chi^2$ of $\sim$3.5 to 5-7 and declines gradually after the maximum and flattens out beyond the reduced $\chi^2$ of ten. About 11\% of all the minimized solutions have reduced $\chi^2$ less than 6.5. These solutions are shown in Fig. \ref{figure:ff3} as a function of the two elliptical frequencies. 
The solutions have been colour-coded to show increasing values of the reduced $\chi^2$ from blue to red. The figure is an expanded view of the parameter space focused on the region where most of the solutions were concentrated.
We added an additional constraint to exclude the parabolic and/or hyperbolic solutions, i.e. where either $e_3$ or $e_4$ are $\geq$ 1.0. 
The overall distribution of the reduced $\chi^2$ as a function of the elliptical frequencies form an arc with chi-square increasing from the lower right through the centre to the upper right regions of Fig. \ref{figure:ff3}. 
The grey star towards the centre of Fig. \ref{figure:ff3} marks the location of the solution presented in \Potters and lies in the region where the intermediate solutions are found.

\subsection{Distribution of the eccentricities}

\begin{figure*}
\centering
\includegraphics[height = 11.00cm, width = 13.0 cm  ]{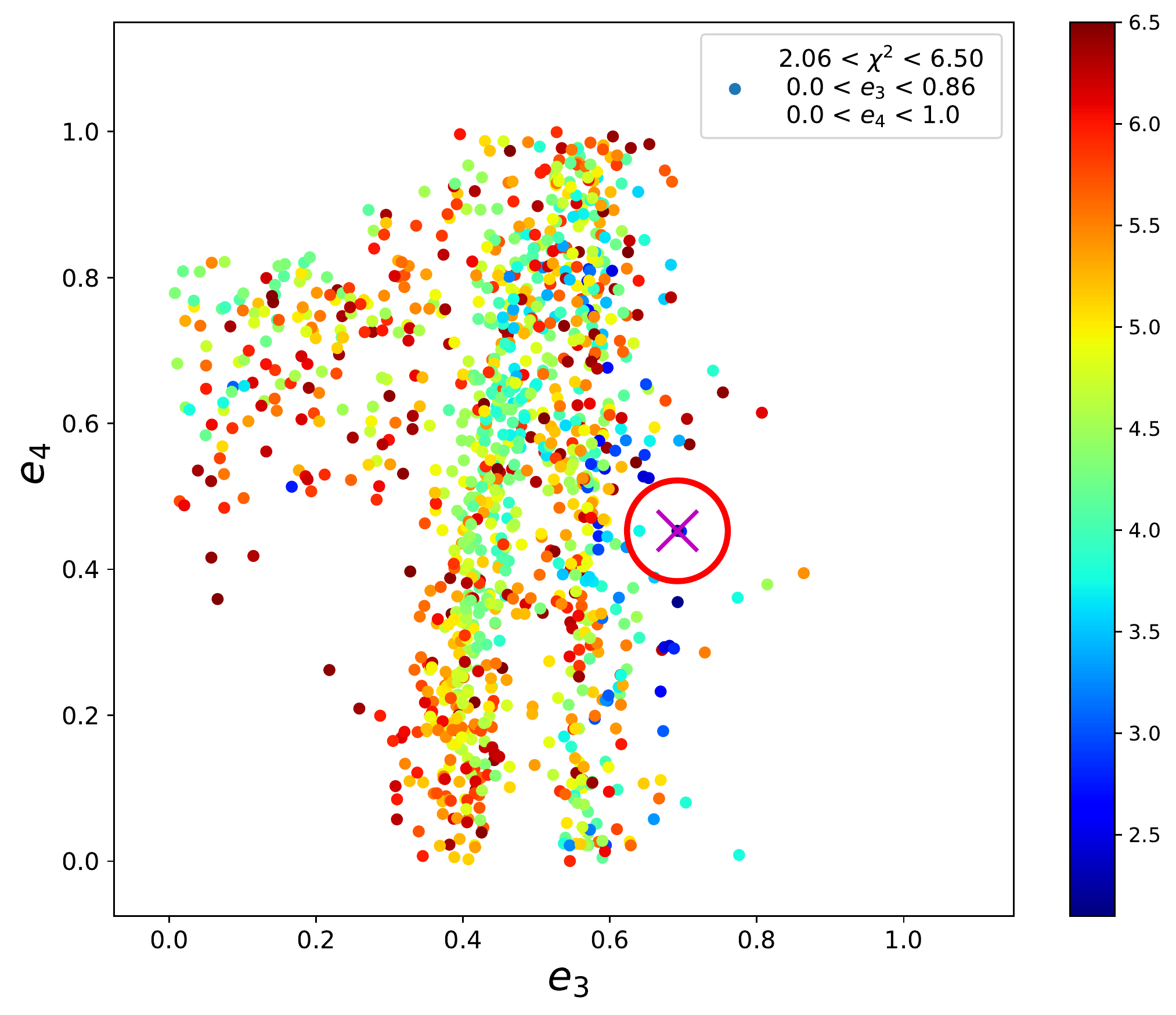}
\caption{Reduced $\chi^2$ parameter space for the two elliptical eccentricities. The points have been colour-coded with dark blue and dark red representing lower and higher reduced chi-square:  2.06 $\leq \chi^2 \leq$ 6.50. The solutions with $e \geq$ 1.0 were excluded since their solutions will give parabolic and hyperbolic orbits. \textit{The red circle and magenta cross mark the solution adopted as the best-fit for this paper.} }
\label{figure:eccen}
\end{figure*}

Figure \ref{figure:eccen} shows the range of eccentricities for the solutions presented in Fig. \ref{figure:ff3}, where $e_3$ and $e_4$ are the eccentricities of the first and second elliptical terms, respectively. 
The solutions have been colour-coded to show increasing values of the reduced $\chi^2$ from blue to red. 
The solutions with the lowest reduced $\chi^2$ values (dark blue circles) form a vertical ridge (first ridge) centred on $e_3$ between $\sim$0.5-0.7 with a wide range of $e_4$ eccentricities between $\sim$0.0-0.9. 
A significant number of solutions form another vertical ridge (second ridge) centred on $e_3 \approx$0.4 and spanning from 0 to 1 for $e_4$. This is associated with the large concentration of light blue and orange to dark red solutions shown in the top right region of Fig. \ref{figure:ff3}. 
The scattered solutions between the two vertical ridges correspond to the lower density of blue and red circles towards the centre of  Fig. \ref{figure:ff3}. 
The solutions to the upper left of the second vertical ridge ($e_3$ < 0.3 and $e_4$ > 0.2) are associated with the solutions with lower $f_4$ frequencies and not shown in Fig. \ref{figure:ff3}.   
Immediately obvious is the absence of solutions in the lower left quadrant corresponding to low values of $e_3$ and $e_4$. There is also a cut-off in the $e_3$ eccentricities at 0.85. 

\subsection{ O - C diagram}

\begin{figure*}
\begin{center}
\includegraphics[height = 13.00cm, width = \textwidth  ]{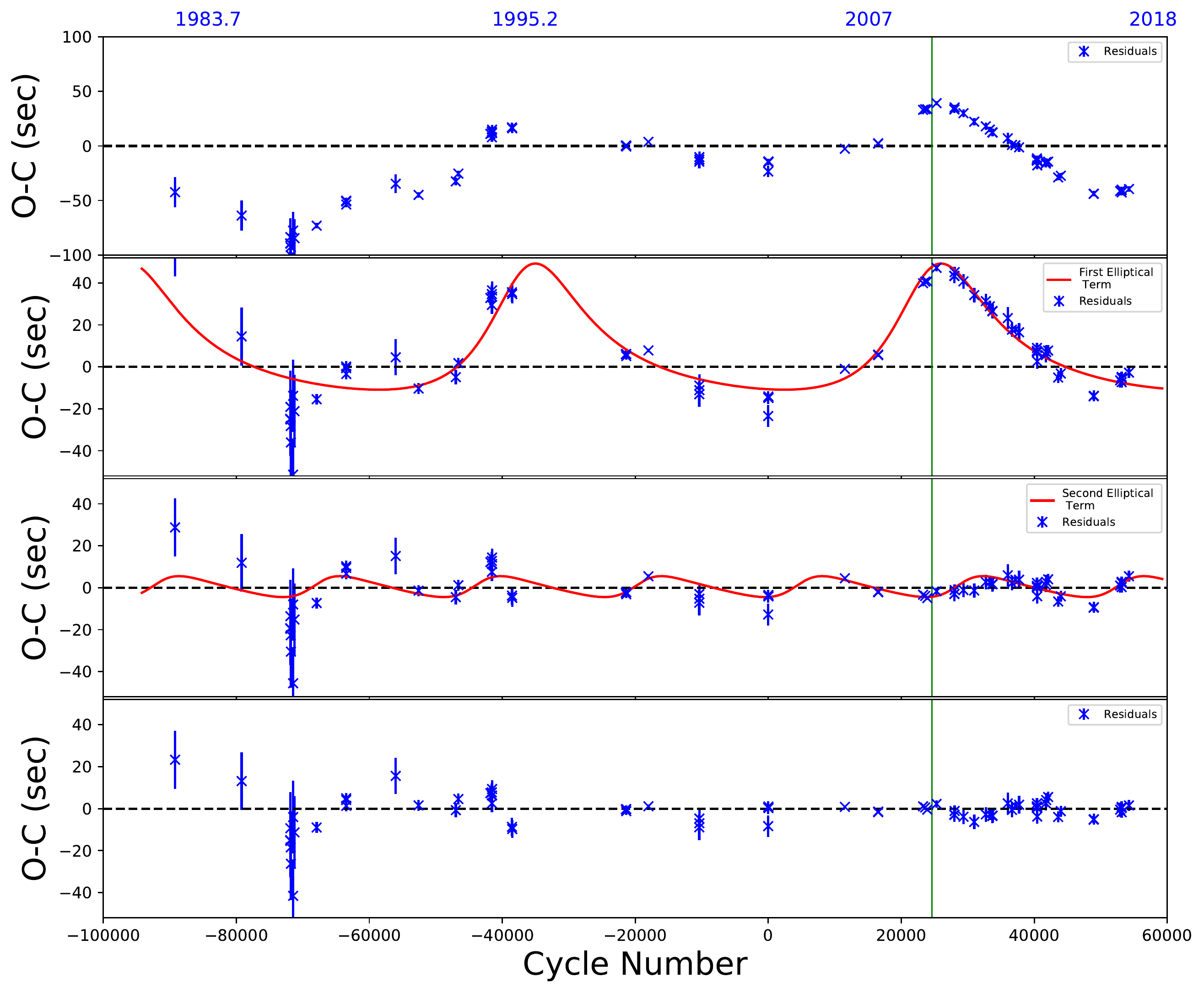}
\end{center}
\caption{Formal best O-C diagram of UZ For, based on new parameters shown in Table \ref{tab:02}, after successive subtraction of the three terms comprising our new eclipse ephemeris. Top: O - C after subtraction of the linear term. Second: O - C after subtraction of the quadratic term with the first elliptical term overplotted (solid red curve). Third: O - C after subtraction of the first elliptical term with the second elliptical term overplotted 
(solid red curve). Bottom: the final O - C residuals after subtraction of the second elliptical term. The vertical green line separate the eclipse times from literature (to the left) and our new eclipse times (to the right). }
\label{figure:oc}
\end{figure*}

Figure \ref{figure:oc} shows the formal best O - C using the solution indicated by a magenta cross and enclosed with a red  circle in both Figs \ref{figure:ff3} and \ref{figure:eccen}. 
This is the solution with the lowest reduced $\chi^2$ (2.06) and is the best-fitting solution. However, there are many other solutions with reduced $\chi^2$ values comparable to this one and indicated by the blue circles in Fig. \ref{figure:eccen}. The parameter values of the best solution are listed in Table \ref{tab:02}. This particular solution has the $e_4$ eccentricity of $\sim$0.45. Similar to the other best solutions, it has an eccentricity of $e_3 \sim$0.69. 
We conclude that the dataset is not consistent with low eccentricities for the two elliptical terms. The best solutions at minimum require the $e_3$ eccentricity to be $\sim$0.7. The residuals in Fig. \ref{figure:oc} are suggestive that by adding a third or fourth elliptical term would reduce the $\chi^2$ further. However, the dataset is currently under-constrained to warrant adding additional terms. We note that in \Potter, they concluded that the best mathematical solution also required high eccentricities. 

\section{Discussion}\label{sec:disc}

In \Potter, they detected departures in the eclipse times of UZ For from a simple quadratic ephemeris of up to $\sim$60 s. They found that the departures also suggest the presence of two elliptical terms with periodicities of $\sim$16(3) and $\sim$5.25(25) years. Similarly, our new results suggest that the deviations in the eclipse O - C shown in Fig. \ref{figure:oc} continue to be best described with a combination of a quadratic term plus two elliptical terms with periodicities of $\sim$14.67(1) and $\sim$5.82(3) years. The deviations are suggestive of both secular and/or periodic variations. Generally, period changes in binary systems are understood as being due to gravitational radiation, magnetic braking, the Applegate mechanism, and the presence of circumbinary planets in orbit around the binary. 

In UZ For, \Potter, they reasoned that the two favoured mechanisms to derive the periodicities are either two giant extrasolar planets as companions to the binary or the Applegate mechanism \citep{1992ApJ...385..621A} due to magnetic cycle activity of the secondary star. 
However, in \Potter, they argued that Applegate's mechanism would require the entire radiant energy output of the secondary and they ruled this mechanism out and recommended a revision such as those described by \cite{1998MNRAS.296..893L}. 
They also argued that a two-planet model was problematic given the quality of the data and a high eccentric orbit for one of the two planets was required in order to capture most of the eclipse times. 
In light of our new results and recent work in the literature we next discuss the two favoured mechanisms. 

\subsection{Circumbinary planets}

Repeating the calculations of \Potters using the solution above, we used the amplitudes of oscillations ($K_{{\rm{bin}},(3,4)}$) to calculate the projected distances $a$\,sin\,$(i)$ from the centre of mass of the binary to the centre of mass of each of the triple systems. 
The centres of mass were 0.064(1) and 0.011(1) au for the long and short periods, respectively. 
Setting the binary mass to be 0.84 $M_{\bigodot}$ (i.e. 0.7 $M_{\bigodot}$ + 0.14 $M_{\bigodot}$, the total combined mass of the WD and red dwarf) gives the corresponding mass functions $f(m_{3,4})$ = 1.2116(3) $\times 10^{-6}$ and 4.045(7) $\times 10^{-8}$ $M_{\bigodot}$. We used the binary inclination ($i$) of $80^{\circ}$ and found the respective minimum masses of the third and fourth body to be 0.00307(5) and 0.00955(2)  $M_{\bigodot}$ and would therefore qualify as extrasolar giant planets [3.22(5) and 10.01(2) $M_J$]. The quoted errors are 1$\sigma$ errors propagated based on the one solution presented and excludes the range in periods shown in Fig. \ref{figure:ff3}. These parameters are summarized in Table \ref{tab:02}.   
 
We note that these two-planet parameters are specific to this one best-fitting solution only. Nevertheless, almost all of the solutions with similar reduced $\chi^2$ gave similar values for the quadratic parameter  $\dot{P}_{{\rm{bin}}}$ = $\frac{2A}{\dot{P}_{{\rm{bin}}}}$ = -3.29(3)$\times$ $10^{-12}$ $\rm s\,s^{-1}$. This corresponds to a rate of angular momentum loss of $\sim 8.3 \times 10^{34}$ erg using equation (5) of \cite{2006MNRAS.365..287B}. The expected theoretical rate of angular momentum loss due to gravitational radiation (Equ. (12) of \citealt{2006MNRAS.365..287B}, corrected from Equ. (2) of \citealt{2003ApJ...582..358A}) and magnetic braking (Equ. (A6) of \citealt{2011ApJS..194...28K} (based on the magnetic braking law by  \citealt{1981A&A...100L...7V})) amounts to $\sim 8.9\times 10^{33}$ erg. Therefore the implied rate of angular momentum loss using the $\dot{P}_{{\rm{bin}}}$ term only,  is ten times larger than the rates of angular momentum loss predicted by gravitational radiation plus magnetic braking alone. 

Within errors, the periodic variations of the formal best solution could be consistent with two planets locked in a 3:1 resonance with orbital periods of 14.67(1) and 5.82(3) years, respectively. However the high eccentricity for both planets implies that such a two planet solution would not be stable. There are solutions of similar reduced $\chi^2$ in which the inner planet has a low eccentricity. Nevertheless all the best solutions require a high eccentricity for the outer planet. Perhaps indicating a one planet solution (see below). 

We certainly need an independent observational approach to shed some light on the existence of planets around UZ For. One such technique may be astrometric monitoring of the precession of the UZ For binary as it wobbles back and forth due to extrasolar companions. The ongoing $GAIA$ mission provides micro arc-seconds parallaxes to thousands of object and place UZ For at a distance of $\sim$240 pc and it might be possible to detect its companion after five years of observations. 
A simulation by \citet{2015MNRAS.448.1118G} suggests that it should be possible to detect $\sim$30 micro arc-seconds for a 7 $M_J$ planet in a 5 au orbit. They argued that for HU Aqr, a polar which shows variations in its O - C diagram and lies at a distance of $\sim$200 pc, it should be possible to detect the outermost companion using parallaxes.  

\subsection{Revised Applegate}

Significant residuals remains in the O - C of UZ For that points to something more complicated than just the presence of two companions to the binary. In \Potter, they noted that the Applegate mechanism would require more than the radiant energy output of the secondary to drive the period changes but suggested that this mechanism is unlikely to be causing the period changes in UZ For. 
\cite{2016A&A...587A..34V} revision of Applegate's mechanism which includes angular momentum exchange between the finite shell and the core of the secondary star place UZ For amongst four systems that could be explained by magnetic activity of the secondary star. 
We conclude that both extrasolar planets and some form of Applegate's mechanism should be considered when explaining the O - C diagram of UZ For. 
Given the smaller amplitude of the residuals after subtraction of the first elliptical term, corresponding to one highly eccentric extrasolar planet, the remaining residuals could be explained by some sort of Applegate mechanism.    

\subsection{Other eclipsing mCVs}

UZ For is not the only post-common-envelope binary which presumably host single- or multiple planetary systems, e.g. HU Aqr \citep{2011MNRAS.414L..16Q,2011MNRAS.416L..11H,2012MNRAS.419.3258W,2015MNRAS.448.1118G} and DP Leo \citep{2010ApJ...708L..66Q,2011A&A...526A..53B}, are other magnetic CVs to show this behaviour. 
For HU Aqr, \citet{2011MNRAS.414L..16Q} reported that eccentricity of the outer planet as big as 0.5, and assumed circular orbit for the inner planet.  
However, revision by \cite{2015MNRAS.448.1118G} suggests eccentricities ranging from 0.1-0.3 for both the inner and the outer planets. 
They also suggest that for stable orbits to exist in the HU Aqr system there must be a third companion orbiting with a very long orbital period and with the middle planet in retrograde orbit. 

\citet{2016MNRAS.460.3873B} found that for the non-magnetic CV, NN Ser, the remaining model has eccentricities ranging from 0.1-0.2 and constrained the period ratio to 2:1 resonance. If both Applegate's mechanism and circumbinary planets are causing the period change in UZ For, this will complicate the process of modelling the eclipse times, since we do not understand both these mechanisms well and it is especially difficult to model the effect of the magnetic activity cycle. We can not say for sure that alternative models such as those of \cite{1992ApJ...385..621A} or angular momentum loss due to gravitational radiation and magnetic braking are not operating on UZ For. 
\cite{2016MNRAS.460.3873B} had suggested that some form of Applegate's mechanism might be at work on NN Ser given its M4 spectral type of the secondary.   

\section{Summary and conclusion}\label{sect:con}

In this paper, we presented and analysed new photometric observations of UZ For together with historical observations collected from literature. We used the new observations to test the two-planet model proposed in \Potters to explain the variations in its eclipse times. 
UZ For undergoes a change in mass transfer from faint to high state, this is captured in Fig. \ref{figure:phot}. The light curves of UZ For show variations in eclipse profiles from one epoch to the next which is consistent with what has been reported in the literature. The eclipse widths remain unchanged and various stages are revealed depending on the accretion state. 

We find that the new mid-eclipse times follow the general trend predicted in \Potters (Fig. \ref{figure:oc_old}) but continues to diverge. In order to accommodate most of the eclipse times, we have recalculated the fitting parameters, including the new data, in the similar manner as in \Potter. We adopted one of the solutions with the lowest reduced $\chi^2$ value as the best fit solution. The parameters for this solution are shown in Table \ref{tab:02} and overplotted in the top and middle panels of Fig. \ref{figure:oc}. The proposed model of the two planets requires the outer planet to have a relatively high eccentric orbit, i.e. $e_{3} =$ 0.69.   
Significant residuals remain as indicated  in the bottom panel of Fig. \ref{figure:oc}.  Adding more elliptical terms (effectively adding more planets) would obviously lower the reduced $\chi^2$ value, however the data is of insufficient quantity to warrant this.

Within errors, the departures in the O - C diagram (Fig. \ref{figure:oc}) are still consistent with the two cyclic variations (14.67(1) and 5.82(3) years) reported in \Potter. However a relatively large eccentricity is required for the longer period and, in addition, seemingly random residuals still remain. 
This suggests that either the circumbinary planet solution is incorrect or requires extra planets, or some form of cyclic magnetic activity is contributing an extra quasi-periodic term to the O - C variations. Further monitoring of the eclipse times is recommended. In the the next 5-10 years, the \textit{GAIA} space mission may be able to detect parallax variations that would be consistent with circumbinary planet solutions as also been suggested by \cite{2015MNRAS.448.1118G} for HU Aqr.



\section*{Acknowledgements}

We would like to thank the anonymous referee whose comments were helpful and improved our manuscript. This material is based upon work supported financially by the National Research Foundation.


\bibliographystyle{aa}
\bibliography{References}

\input{Table_mid_eclipse}

\end{document}

%% file: Phot_obsev_log.tex
\begin{table*}
\caption{Observational log of UZ For. All observations were made with the HIPPO and SHOC instruments on the SAAO 1.0-m and 1.9-m telescope. }
\centering
\label{Table:phot}
\begin{tabular}{ l l l c c c }\hline \hline
 Date of & Telescope & Instrument  & Length of & Binning & Number of  \\
 Observation & SAAO & used & observations & &  eclipse(s)  \\
 &&& (hours) & &  \\ \hline \hline
 2018-02-21  & 1.9-m & HIPPO  & 1.00 & - & 1  \\
 2017-11-16  & 1.9-m & HIPPO  & 5.12 & - & 3  \\
 2017-10-27  & 1.0-m & SHOC   & 2.25 & 8 $\times$ 8 & 1  \\
 2016-11-14  & 1.0-m & SHOC   & 2.03 & 8 $\times$ 8 & 1  \\
 2016-11-11  & 1.0-m & SHOC   & 2.08 & 8 $\times$ 8 & 1  \\
 2015-09-04  & 1.9-m & SHOC   & 1.00 & 8 $\times$ 8 & 1  \\
 2015-07-31  & 1.9-m & SHOC   & 1.00 & 8 $\times$ 8 & 1 \\
 2015-03-21  & 1.9-m & HIPPO  & 1.35 & - & 1 \\
 2015-02-23  & 1.0-m & SHOC   & 0.85 & 16 $\times$ 16 & 1 \\
 2015-02-18  & 1.0-m & SHOC   & 2.06 & 16 $\times$ 16 & 1 \\ 
 2014-10-28  & 1.9-m & SHOC   & 2.70 & 16 $\times$ 16 & 2 \\
 2014-10-26  & 1.9-m & SHOC   & 0.62 & 16 $\times$ 16 & 1 \\
 2014-10-24  & 1.9-m & SHOC   & 0.77 & 16 $\times$ 16 & 1 \\
 2014-10-23  & 1.9-m & SHOC   & 1.17 & 16 $\times$ 16 & 1 \\
 2014-10-22  & 1.9-m & SHOC   & 1.00 & 16 $\times$ 16 & 1 \\
 2014-03-02  & 1.9-m & HIPPO  & 1.65 & - & 1 \\
 2014-01-12  & 1.9-m & SHOC   & 1.13 & 8 $\times$ 8 & 1 \\
 2013-11-28  & 1.9-m & HIPPO  & 1.68 & - & 1 \\
 2013-10-03  & 1.0-m & SHOC   & 0.52 & 16 $\times$ 16 & 1 \\
 2013-03-17  & 1.9-m & HIPPO  & 1.11 & - & 1 \\
 2013-03-16  & 1.9-m & HIPPO  & 0.79 & - & 1 \\ 
 2013-03-13  & 1.9-m & HIPPO  & 1.43 & - & 1 \\
 2013-02-08  & 1.9-m & SHOC   & 1.02 & 8 $\times$ 8 & 1 \\
 2012-12-17  & 1.0-m & SHOC   & 0.59 & 16 $\times$ 16 & 1 \\
 2012-07-17  & 1.9-m & HIPPO  & 0.88 & - & 1 \\
 2012-02-26  & 1.9-m & HIPPO  & 1.63 & - & 1 \\
 2011-11-01  & 1.9-m & HIPPO  & 1.26 & - & 1 \\
 2011-10-31  & 1.9-m & HIPPO  & 0.90 & - & 1 \\
 2011-10-27  & 1.9-m & HIPPO  & 2.60 & - & 1 \\
 2011-03-08  & 1.9-m & HIPPO  & 0.37 & - & 1 \\  \hline \hfill 
\end{tabular}
\end{table*}

%% file: Table_mid_eclipse.tex
\onecolumn

\begin{table*}
\caption{Mid-eclipse ephemeris of the main accretion spot of UZ For and corresponding planet model parameters. The ephemeris are rounded off to the 
1$\sigma$ errors. The planet parameter errors are 1$\sigma$ errors and were propagated from the one fitting solution and may underestimate true errors of range in parameter space of possible solutions. The minimum planet masses are listed assuming coplanearity and $M_{3,4,fnc}$ is the mass function. The combined mass of the primary and secondary stars is assumed to be 0.84$\rm{M_{\odot}}$. (Table reproduced from--\citealt{2011MNRAS.416.2202P}).}
\label{tab:02}
\centering
\begin{tabular}{l l l }\hline
Quadratic term & $T_0$ = 2453405.300833(5) d & \\
               & $P_{bin}$ = 0.087865421(1) d & Planet \\
               & $A$ = -14.5(2)$\times 10^{-14}$  & Parameters \\ \\
1st elliptical term& $\upsilon_3$ = ($E + T_3$)$f_3$ & $M_{3,fnc}$ = 1.326(7)$10^{-6}\rm{M_{\odot}}$ \\ 
               &$T_3$ = 67198(145) (binary cycle)     & $M_{3,Jup}$ = 10.00(2) \\
               &$f_3$ = 0.0001030(1) (cycles per binary cycle) & $P_3$ = 14.67(1) years\\
               &$\varpi_3$ = 2.10(1)                 & $a_3$ = 5.7(1) au \\
               &$K_{bin,(3)}$ = 0.000371(3) d & $a_{1,3}$ = 0.064(1) au \\
               & $e_3$ = 0.69(1) & \\ \\
2nd elliptical term & $\upsilon_4$ = ($E + T_4$)$f_4$ & $M_{4,fnc}$ = 3.43(9)$10^{-8}\rm{M_{\odot}}$ \\ 
               &$T_4$ = 7444(219) (binary cycle)       & $M_{4,Jup}$ = 3.22(5) \\
               &$f_4$ = 0.000260(1) (cycles per binary cycle) & $P_4$ = 5.82(3) years\\
               &$\varpi_4$ = -0.22(5)                  & $a_4$ = 3.0(2) au \\
               &$K_{bin,(4)}$ = -0.000065(3) d & $a_{1,4}$ = 0.011(1) au \\
               & $e_4$ = 0.45(6)  & \\ \\ \hline
\end{tabular}
\end{table*}  

\begin{longtable}{r l c c c c c}
\label{tab:01}\\
\caption{Mid-eclipse times of the main accretion spot of UZ For. $\rm{BJD_{TDB}}$ is the BJD in the TDB system. The ingress and egress times have the integer of BJD subtracted. All the times have been barycentrically corrected. }\\ \hline
Cycle &$\rm{BJD_{TDB}}$+2400000& $\Delta$$\rm{BJD_{TDB}}$ & Width (s) & $\rm{T_{ingress}}$ & $\rm{T_{egress}}$ & Reference \\ \hline
\hline
\endfirsthead
\multicolumn{4}{c}%
{\tablename\ \thetable\ -- \textit{Continued from previous page}} \\
\hline
Cycle &$\rm{BJD_{TDB}}$+2400000 & $\Delta$$\rm{BJD_{TDB}}$ & Width (s) & $\rm{T_{ingress}}$ & $\rm{T_{egress}}$ & Reference \\
\hline
\endhead
\hline \multicolumn{7}{r}{\textit{Continued on next page}} \\
\endfoot
\hline
\endlastfoot
54242 & 58171.29654185  & 0.00003 & 472(3) & 0.293811(20) & 0.299227(20) & $^{znk}$ \\ 
53141 & 58074.55667742  & 0.00003 & 471(3) & 0.553950(20) & 0.559405(20) & $^{znk}$ \\ 
53140 & 58074.46884132  & 0.00003 & 472(2) & 0.466109(20) & 0.471574(20) & $^{znk}$ \\ 
53139 & 58074.38097010  & 0.00002 & 471(2) & 0.378244(10) & 0.383696(10) & $^{znk}$ \\ 
52912 & 58054.43550982  & 0.00004 & 471(4) & 0.432808(50) & 0.438234(60) & $^{znk}$ \\ 
48962 & 57707.36706944  & 0.00003 & 471(2) & 0.364344(20) & 0.369795(20) & $^{znk}$ \\  
48928 & 57704.37964299  & 0.00002 & 472(2) & 0.376910(10) & 0.382376(10) & $^{znk}$ \\  
43991 & 57270.58825176  &  0.00003 & 471(3) & 0.585528(20) & 0.590976(20) & $^{znk}$ \\  
43593 & 57235.61779679  &  0.00004 & 470(3) & 0.615071(30) & 0.620522(20) & $^{znk}$ \\
42087 & 57103.29264038  &  0.00003 & 472(3) & 0.289908(30) & 0.295372(20) & $^{znk}$ \\
41791 & 57077.28447068  &  0.00004 & 472(4) & 0.281738(30) & 0.287203(30) & $^{znk}$ \\  
41735 & 57072.36399146  &  0.00004 & 472(3) & 0.361262(30) & 0.366721(30) & $^{znk}$ \\  
40450 & 56959.45690562  &  0.00004 & 470(3) & 0.454187(50) & 0.459624(80) & $^{znk}$ \\
40449 & 56959.36909531  &  0.00002 & 471(1) & 0.366371(20) & 0.371819(30) & $^{znk}$ \\
40427 & 56957.43605536  &  0.00002 & 471(1) & 0.433331(30) & 0.438779(30) & $^{znk}$ \\
40404 & 56955.41517125  &  0.00003 & 472(2) & 0.412439(30) & 0.417903(40) & $^{znk}$ \\ 
40393 & 56954.44862704  &  0.00002 & 472(1) & 0.445895(20) & 0.451359(30) & $^{znk}$ \\
40381 & 56953.39425098  &  0.00002 & 471(1) & 0.391527(40) & 0.396975(20) & $^{znk}$ \\
37717 & 56719.32090166  &  0.00006 & 470(3) & 0.317920(80) & 0.323621(30) & $^{znk}$ \\ 
37160 & 56670.37988451  &  0.00002 & 471(1) & 0.377161(20) & 0.382608(40) & $^{znk}$ \\
36648 & 56625.39279285  &  0.00004 & 471(3) & 0.390068(30) & 0.395518(30) & $^{znk}$ \\
36013 & 56569.59832086  &  0.00006 & 475(5) & 0.595571(40) & 0.601070(40) & $^{znk}$ \\  
33733 & 56369.26522176  &  0.00004 & 472(3) & 0.262493(50) & 0.267951(60) & $^{znk}$ \\
33722 & 56368.29859146  &  0.00004 & 471(1) & 0.295865(20) & 0.301318(30) & $^{znk}$ \\
33688 & 56365.31135803  &  0.00004 & 471(2) & 0.308634(60) & 0.314082(40) & $^{znk}$ \\
33313 & 56332.36177629  &  0.00003 & 470(2) & 0.359056(20) & 0.364497(20) & $^{znk}$ \\  
32710 & 56279.37896176  &  0.00004 & 472(3) & 0.376250(30) & 0.381696(30) & $^{znk}$ \\  
30972 & 56126.66890920  &  0.00004 & 470(2) & 0.666182(60) & 0.671642(50) & $^{znk}$ \\
29352 & 55984.32701723  &  0.00004 & 470(1) & 0.324295(40) & 0.329740(20) & $^{znk}$ \\
28023 & 55867.55393431  &  0.00003 & 472(2) & 0.551204(50) & 0.556664(40) & $^{znk}$ \\
28010 & 55866.41166271  &  0.00002 & 471(2) & 0.408936(40) & 0.414390(30) & $^{znk}$ \\
27966 & 55862.54558562  &  0.00004 & 469(3) & 0.542871(60) & 0.548301(60) & $^{znk}$ \\ 
25311 & 55629.26295762  &  0.00002 & 470(1) & 0.260238(20) & 0.265677(30) & $^{znk}$ \\
23913  & 55506.42703435& 0.00001   & 468(2) && & $^a$\\
23595  & 55478.48583116& 0.00001   & 468(2) && & $^a$\\
23277  & 55450.54462082& 0.00001   & 467(2) && & $^a$\\
16526  & 54857.36480850& 0.00001   & 469(2) && & $^a$\\
16526  & 54857.36480517& 0.0000086 & 469(1) && & $^a$\\
11518  & 54417.33472170& 0.0000086 & 468(1) && & $^a$\\
34     & 53408.28808581& 0.0000086 & 469(1) && & $^a$\\
23     & 53407.32157438& 0.00001   & 469(2) &&  & $^a$\\
0      & 53405.30066303& 0.000035  & 469(3) &&  & $^a$\\
-11    & 53404.33404192& 0.00006   & 467(4) &&  & $^a$ \\
-10362 & 52494.83919610& 0.000087  & 479(8) &&  & $^a$ \\
-10365 & 52494.57562568& 0.000035  & 469(3) &&  & $^a$\\
-10376 & 52493.60905802& 0.00007   & 469(6) &&  & $^a$\\
-18023 & 51821.70239393& 0.00001   & 467(2) &&  & $^b$\\
-21360 & 51528.49543399& 0.00002   & 468(2) &&  & $^c$\\
-21361 & 51528.40757990& 0.00002   & 468(2) &&  & $^c$\\
-21429 & 51522.43272958& 0.00002   & 468(2) &&  & $^c$\\
-38508 & 50021.77938800& 0.00005   & &&  & $^d$\\
-38543 & 50018.70410800& 0.00005   & &&  & $^d$\\
-41537 & 49755.63497800& 0.00005   & &&  & $^d$\\
-41538 & 49755.54714800& 0.00005   & &&  & $^d$\\
-41560 & 49753.61402800& 0.00005   & &&  & $^d$\\
-41571 & 49752.64756800& 0.00005   & &&  & $^d$\\
-41790 & 49733.40501704& 0.00004   & 467(4) && & $^a$\\
-46605 & 49310.33259382& 0.00003   & 471(4) && & $^a$\\
-46988 & 49276.68005500& 0.00004   &        && & $^e$\\
-52587 & 48784.72141928& 0.00003   & 463(4) && & $^a$\\
-56024 & 48482.72808573& 0.0001    & 477(5) && & $^f$\\
-63462 & 47829.18486375& 0.00003   &        && & $^g$\\
-63474 & 47828.13052000& 0.00003   &&& & $^g$\\ 
-63476 & 47827.95478000& 0.00003   &&& & $^g$\\
-67915 & 47437.91992000& 0.00003   & 466.5(2.5)&& & $^{f,h}$\\
-71248 & 47145.06433900& 0.0002    &&& & $^i$\\
-71451 & 47127.22773900& 0.0002    &&& & $^i$\\
-71452 & 47127.13943900& 0.0002    &&& & $^i$\\
-71786 & 47097.79255900& 0.0002    &&& & $^j$\\
-71821 & 47094.71735900& 0.0002    &&& & $^j$\\
-71857 & 47091.55423900& 0.0002    &&& & $^j$\\
-71868 & 47090.58778900& 0.0002    &&& & $^j$\\
-71889 & 47088.74254900& 0.0002    &&& & $^j$\\
-79193 & 46446.97380900& 0.00016   &&& & $^k$\\
-89206 & 45567.17759700& 0.00016   &&& & $^k$\\ \hline 
\end{longtable}
\begin{flushleft}
$^a$\cite{2011MNRAS.416.2202P}; $^b$\cite{2002OptEn..41.1158D}; $^c$\cite{2001MNRAS.324..899P}; $^d$\cite{1998ApJ...501..830I}; 
$^e$\cite{1995ApJ...445..909W}; $^f$\cite{1994IBVS.4075....1R}; $^g$\cite{1991MNRAS.253...27B}; $^h$\cite{1989ApJ...347..426A}; 
$^i$\cite{1989ApJ...337..832F}; $^j$\cite{1988A&A...195L..15B}; $^k$\cite{1988ApJ...328L..45O}; $^{znk}$This Paper.
\end{flushleft}